# Observation of tungsten impurity suppression with on-axis ECRH by an X-ray Crystal Spectroscopy on EAST

Z.C. Lin[1,2], H.M. Zhang [1]*, F.D. Wang[1]*, Cheonho Bae[1], J. Fu[1], Y. C. Shen[3], S.Y. Dai[4], D.A. Lu[1,2], Y.F. Jin[1,2], L. He[1,2], M.R. Wang[1,2], G.L. Lin[1,2], K.X. Ye[1], S.X. Wang[1], H.L. Zhao[1], B. Lyu[1,2]

[1]*Institute of Plasma Physics, HFIPS, Chinese Academy of Sciences, Hefei 230031, China*

[2]*Science Island Branch, Graduate School of University of Science and Technology of China, Hefei, 230026, China*

[3]*School of Physics and Materials Engineering, Hefei Normal University, Hefei 230601, China*

[4]*Key Laboratory of Materials Modification by Laser, Ion and Electron Beams (Ministry of Education), School of Physics, Dalian University of Technology, Dalian 116024, China*

hmzhang@ipp.ac.cn; fdwang@ipp.ac.cn;

**Abstract**

Impurity degrades tokamak plasmas confinement by causing energy loss, diluting the fuel concentration, even terminating the discharges in some extreme cases. Previously, the suppression effects of on-axis Electron Cyclotron Resonance Heating (ECRH) on the impurity accumulation have been investigated on EAST by the extreme ultraviolet (EUV) spectroscopy. However, it is difficult to quantify the changes in impurity tungsten (W) profile since the W line emissions in the EUV range could not be easily resolved. The X-ray Crystal Spectroscopy (XCS), that used to provide the ion temperature and the rotation velocity by measuring lines emissions in the soft X-ray range, also can be used to study the behavior of impurity W emissions. To begin with, in-situ absolute intensity calibration for Tangential XCS (TXCS) is conducted by analyzing the measurements of the bremsstrahlung radiation intensity. After obtaining the calibration coefficient, $W^{44+}$ ion density profiles are evaluated by Abel inversion using the spectral line of W XLV (3.9095 Å). Thus, a direct observation of $W^{44+}$ impurity concentration suppressed by ECRH is accomplished. The obtained W density profiles can be used to analyze the W transport by combining with the impurity transport codes in the future.

## I. Introduction

Impurity accumulations in the core region can cool down the tokamak plasmas, enhance the radiation loss, and dilute the density of fuel particles.[1] Suppressing the impurity accumulation is one of the critical issues for achieving high-performance long-pulse operations in tokamaks. Due to the high heat flux from the main plasmas, tungsten (W) with high melting point, is usually selected as the plasma facing material. Therefore, the measurement of impurity W density profile, can provide a direct method to confirm whether impurity accumulation is suppressed. Previously, the measurements of impurity molybdenum (Mo) profiles by the extreme ultraviolet (EUV) spectroscopy have been used to analyze the suppression effects of Electron Cyclotron Resonance Heating (ECRH) on EAST. However, it is still difficult to obtain the W density profiles in the EUV range due to the lack of well-resolved W spectral lines.[2,3]

The X-ray crystal spectroscopy (XCS) as a key diagnostic to obtain the ion temperature and rotation velocity has been developed on EAST. Recently, a resolved spectral line of $W^{44+}$ is identified at the wavelength of 3.9095 Å, which provides the possibility to obtain the W ion density profile using the XCS.[4] Given the lack of direct observation of W density in the EUV range and the recently discovered W spectra with the XCS, this paper proposes a new method to measure the W profiles using the intensity-calibrated XCS system on EAST.

Since the absolute intensity calibration of spectroscopy system is the prerequisite to calculate the impurity density profiles, EUV spectroscopies on EAST and HL-2A have already been well absolutely calibrated on the basis of bremsstrahlung measurement.[5-8] In this study, therefore, the XCS system is also absolutely calibrated in a similar method. Then, non-circular cross-section Abel inversion is conducted to obtain the radial profiles of impurity W. Finally, the impurity suppression effect by on-axis ECRH is directly observed on the basis of W density profile measurements. In this paper, the XCS system on EAST is briefly introduced in Section 2. Method of the absolute intensity calibration and the principle of Abel inversion to obtain impurity emission profiles are shown in Section 3. The impurity suppression effect by the ECRH is presented in Section 4. The study is summarized in Section 5.

## II. XCS system on the EAST

XCS on EAST is composed of the Tangential XCS (TXCS) system and the Poloidal XCS (PXCS) system, and the TXCS is used.[9-13] In this paper, only TXCS is used for analysis. According to the XCS setup shown in Fig. 1, the crystal's radial position of TXCS is $R = 10.95$ m. The crystal with effective area of $80 \times 80$ mm$^2$ is installed in the TXCS crystal chamber. The detector is separated from the crystal by 3.032 m for TXCS on the radial direction. The PILATUS 900K detector, which consists of 9 modules, is used to detect the X-ray spectra. Each module of the detector consists of $487 \times 195$ pixels with each pixel size of $172 \times 172$ μm$^2$. Figure 2 shows the result of spatial position calibrations for TXCS system, illustrating the relationship between the vertical position of detector and the relative position to the equatorial plane on EAST. Figure 3 shows the typical spectra of TXCS, with a number of new identified W spectral lines, e.g., the W XLV (3.9095 Å). As a result, the spectra measurements of TXCS are used to analyze the W impurity behavior in this paper.

The measured chord-integrated emission intensity $P_{line}(\lambda)$ of the TXCS could be expressed as

$$P_{line}(\lambda) = \int_0^1 P_{local}(\lambda,\rho) \, r(\rho) \, \Delta\Omega(\rho) \, d\rho \tag{1}$$

where $P_{local}(\lambda,\rho)$ is the local emissivity, $r(\rho)$ is the length of chord, and $\Delta\Omega(\rho)$ is the solid angle of the TXCS system to collect the plasma radiation. $\Delta\Omega$ can be expressed as follows

$$\Delta\Omega = \frac{\Delta A}{R^2} \tag{2}$$

where $\Delta A$ is the effective area of the TXCS crystal, and $R$ is the distance between the crystal and the plasma. The maximum value of $R$ is $R_{max} = 11.4$ m, while the minimum value is $R_{min} = 10.5$ m for TXCS in EAST. If the solid angle of the TXCS system at magnetic axis is expressed as $\Delta\Omega_0$, the error between maximum and minimum value of the solid angle is

$$\frac{\Delta\Omega_{max}-\Delta\Omega_0}{\Delta\Omega_0} \approx \frac{\Delta\Omega_0-\Delta\Omega_{min}}{\Delta\Omega_0} \approx 0.1\Delta\Omega_0. \quad (3)$$

By calculating with Eq. (2), $\Delta\Omega_0=5.3377\times10^{-5}$ sr is obtained for TXCS in EAST.

## III. Absolute intensity calibration of TXCS

Absolute intensity calibration could be traditionally performed with the help of X-ray standard source or synchrotron radiation light source. However, in this case, it is necessary to install require installation the XCS spectroscopy at the light source. Therefore, in this study an in-situ absolute intensity calibration method is proposed for TXCS. Calibration coefficients are obtained by comparing the calculated bremsstrahlung of EAST plasma to the measured intensity from TXCS.

### A. Bremsstrahlung in the X-ray range

Continuum emissions of plasmas include bremsstrahlung and recombination radiations. However, during the general stable discharges in Tokamak, recombination radiation can be negligible compared with the bremsstrahlung in the X-ray range. [6]Assuming that the velocity distribution of electrons is Maxwellian, the bremsstrahlung is expressed as follows

$$U_{brem} = 1.9\times10^{-34} \frac{n_e^2 Z_{eff}^2 g_{ff}}{\lambda^2 T_e^{1/2}} \times \exp\left(\frac{-12395}{\lambda T_e}\right) \left[\text{W}\cdot\text{m}^{-3}\cdot\text{Å}^{-1}\right] \quad (4)$$

where $n_e$ is the electron density in the unit of m$^{-3}$, $T_e$ is the electron temperature in unit of eV, $Z_{eff}$ is the effective ion charge, $\lambda$ is the wavelength in the unit of Å, and the $g_{ff}$ is the correction factor called gaunt factor. Gaunt factor is related to the electron temperature and the wavelength that is calculated by the RADZ1 code.[14]

### B. The principle of Abel inversion

As mentioned above, the measurements of the XCS system are line-integrated profiles of impurity emssion. However, the radial profiles of impurity emission intensity are more beneficial for physical analysis. Therefore, it is necessary to calculate the radial profiles using the Abel inversion method. Since EAST plasma has a non-circular poloidal cross section, the analytic form of Abel inversion is not applicable. Assuming that the emission intensity is uniform along the magnetic surface $j$, the line-integrated intensity of the chord $i$ can be expressed as follows

$$B_i = \sum_j L_{ij} E_j \qquad (5)$$

where $B_i$ is the intensity of chord $i$, $E_j$ is the local emission intensity of the region $j$, and $L_{ij}$ is the length matrix of their intersection. For convenience, the chords are selected tangent to the magnetic surface, so the $L_{ij}$ could be divided into two parts.

$$L_{ij} = L_{ij}^{HFS} + L_{ij}^{LFS} \qquad (6)$$

where $L_{ij}^{HFS}$ is the distance from the high filed side (HFS) to the point of tangency and the $L_{ij}^{LFS}$ is the distance from the low field side (LFS) to the tangent point as shown in Fig. 4. Therefore, the form of $L_{ij}$ can be transformed into an upper triangular matrix, which has its inverse matrix. The $E_j$ is expressed as the matrix product of the inverse chord length matrix $L_{ij}^{-1}$ and the line-integrated brightness matrix $B_i$ as follows

$$E_j = \sum_i L_{ij}^{-1} B_i. \qquad (7)$$

## C. The absolute intensity calibration based on the bremsstrahlung

The purpose of absolute intensity calibration is to determine the calibration coefficients between the count rate of photon-electrons measured by the detector and the absolute intensity of impurity emissions. The bremsstrahlung of fusion plasmas can be calculated on the basis of several fundamental plasma parameters provided by other diagnostics, i.e., the line-integrated density measured by the POlarimeter-INTferometer (POINT), the $T_e$ measured by the Electron Cyclotron Emission (ECE) or Thomson Scattering (TS), and the $Z_{eff}$ measured by the absolutely calibrated visible

spectroscopy.[15-18]

For the purpose of calculation, $I_{\text{brem\_obs}}$ is used to present the inversed bremsstrahlung intensity (counts·s$^{-1}$·Å$^{-1}$), and $I_{\text{brem\_cal}}$ is the calculated bremsstrahlung intensity (photons·m$^{-3}$·Å$^{-1}$·s$^{-1}$·sr$^{-1}$). Then, the calibration coefficients $F_{cali}$ can be expressed as

$$F_{cali} = \frac{I_{\text{brem\_cal}}}{I_{\text{brem\_obs}}}. \tag{8}$$

In this study, $F_{cali}$ is calculated with the data from several different discharges. Before performing $F_{cali}$ calculations, it is necessary to confirm whether the continuum spectra measured by TXCS are dominantly contribution of bremsstrahlung contributions. According to the expression of bremsstrahlung intensity, the intensity should be proportional to the square of the density $n_e^2$. Figure 5(a) shows the relationship between the inversed continuum spectra intensity $I_{\text{contin}}$ and $n_e^2$ from shot the #103159. It is found that there is a clear linear relationship between $I_{\text{contin}}$ and $n_e^2$. Figure 5(b) shows the relation between $I_{\text{contin}}$ with $n_e$ from similar condition shots whose $T_e$ are $4.9 \pm 0.3$ keV and the value of $Z_{eff}$ in the range between 1.4 and 2. It manifests a clear quadratic relationship between $I_{contin}$ with $n_e$. Therefore, spectra in these wavelength ranges can be used for conducting the absolute intensity calibrations. Figure 5(c) shows the obtained calibration coefficients. It is shown that the coefficients are mostly around the $10^{10}$ photons·m$^{-3}$·Å$^{-1}$·s$^{-1}$·sr$^{-1}$/(counts·s$^{-1}$·Å$^{-1}$). Therefore, the average value of $3.62 \times 10^{10}$ photons·m$^{-3}$·Å$^{-1}$·s$^{-1}$·sr$^{-1}$/(counts·s$^{-1}$·Å$^{-1}$) is used to present $F_{cali}$ in the subsequent calculations.

## D. The measurement of the W$^{44+}$ emission profiles

Difference in the impurity behaviors between low and high $T_e$ discharges are studied by comparing the data of #98958 and #107006 discharges. The basic parameters of these two shots are presented in Fig. 6, while $T_e$ and $n_e$ profiles are shown in Fig.7. Since the position of TXCS detector has slightly moved, an independent calibration using the full profile has been applied for #107006 to evaluate the correctness and accuracy of the previous approach. The $n_e$ is provided by the reflectometer, the $T_e$ is provided by the ECE, and the $Z_{eff}$ is set at 4.5 from visible spectroscopy. Because of the movement of the detector, the wavelengths of the bremsstrahlung have been rechosen. Figure 8 shows an example of aforementioned calibration technique with the full profiles of $n_e$ and $T_e$, calculation profile keeps a very consistent agreement with the measured profile. It is also found that the calculated calibration coefficient for shot #107006 is consistent with the previous results presented in Fig.5. Therefore, the calculation results demonstrate the accuracy and self-consistency of the absolute calibration method applied in this work.

The emission intensity of W XLV (3.9095 Å) can be expressed as

$$E_{W^{44+}}^{0.39095\ nm} = n_Z \cdot n_e \cdot PEC^{exc} + n_{Z+1} \cdot n_e \cdot PEC^{rec} + n_{Z+1} \cdot n_H \cdot PEC^{CX} \tag{9}$$

where the intensity includes the contributions of the excitation, recombination, and charge exchange. Since the two discharges plasmas in Fig.7 are at the steady-state phase without NBI injections, the second and the third terms could be neglected.[3] Thus, the density of $W^{44+}$ can be expressed as follows

$$n_{W^{44+}} = \frac{E_{W^{44+}}^{0.39095\ nm}}{n_e \cdot PEC^{exc}} \tag{10}$$

where $PEC^{exc}$ can be obtained from ADAS database.[19] Since the $PEC^{exc}$ is not sensitive to the density but to the temperature, the interpolation is done according to the $T_e$ profile measured by ECE, and $n_e$ of #98958 is from the POINT. Figure 9 shows the results of the emission and the density profiles. Because $T_e$ of #98958 is above 10 keV which is significantly higher than the 6 keV of #107006, the peak position of the $W^{44+}$ ion density profile is shifted to outer radial location in accordance with the upper energy level of the 3.178 keV of this spectral line. According to the $W^{44+}$ ion density profiles, it is estimated that the $W^{44+}$ density is smaller than electron densities by 4-5 orders of magnitude in typical EAST discharges.

### E. The error analysis of the impurity density profile

The error of the absolute intensity calibration mainly comes from the measurement error of the continuous spectra from TXCS and the calculation error of bremsstrahlung intensity on the basis of the measurements of other diagnostics. The measurement error can be decreased by repeated measurements and the average calculation. The calculation error of impurity density mainly comes from the $T_e$, $n_e$, $Z_{eff}$, and the magnetic surface information from the EFIT.[20]

Assuming that $N$ is a calculated value from the inputs of $x_1$, $x_2$, $x_3$ and $x_4$, the relative error can be expressed as follows from the error propagation theory

$$\frac{S_N}{N} = \sqrt{\left(\frac{\partial \ln f}{\partial x_1}\right)^2 S_{x_1}^2 + \left(\frac{\partial \ln f}{\partial x_2}\right)^2 S_{x_2}^2 + \left(\frac{\partial \ln f}{\partial x_3}\right)^2 S_{x_3}^2 + \left(\frac{\partial \ln f}{\partial x_4}\right)^2 S_{x_4}^2}. \quad (11)$$

According to the bremsstrahlung calculation and the measurement intensity of the TXCS, the relative error is given by

$$\frac{\Delta F_{cali}}{F_{cali}} = \sqrt{\left(\frac{\Delta Z_{eff}}{Z_{eff}}\right)^2 + \left(\frac{2\Delta n_e}{n_e}\right)^2 + \left(\frac{\Delta T_e}{2T_e}\right)^2 + \left(\frac{\Delta r}{r}\right)^2 + \left(\frac{\Delta P_{Expe}}{P_{Expe}}\right)^2}. \quad (12)$$

The relative error of the $T_e$ is ~10%, the $n_e$ is replaced by the line-integrated density that causes a ~20% margin of error, and the relative error of the $Z_{eff}$ is ~20%, the error in the magnetic surface position by EFIT is ~5%, and the error of the experiment measurement and profile fit is about ~20%, which yield the final accumulated error at about 49.5% for the impurity density measurements.

## IV. The measurement of tungsten impurity accumulation suppression by ECRH

Previously, the suppression of the impurity Mo accumulation has been analyzed based on the EUV spectroscopy on EAST.[2,3] However, it is more difficult to measure the W emission with the EUV spectroscopy due to its insufficient spectral resolution. Therefore, the TXCS with enough spectral resolution provides an alternative method to measure the impurity W density as mentioned above. The measured W density profiles by TXCS could be very useful for analyzing impurity behaviors in the core region.

Figure 10 shows the discharge waveforms of EAST #103186. As the ECRH is applied at t = 2.4 s, the emission intensity of $W^{44+}$ decreases rapidly. The impurity W density behaviors are analyzed by comparing the data at two time slices at t = 2.3 s and 4.5 s. Based on the profiles measured by ECE and reflectometer as shown in Fig. 11(a) and (b), the density profiles of the $W^{44+}$ ions are obtained. It is shown that the $W^{44+}$ density drops dramatically after the ECRH injected. Besides, due to the suppression effect of ECRH on impurity accumulation, the $W^{44+}$ density profile at t = 4.5 s becomes flatter. The total averaged density of the $W^{44+}$ also decreases from $2.79 \times 10^{15}$ to $2.04 \times 10^{14}$ m$^{-3}$, i.e., by an order of magnitude, while the intensity of low-Z impurities such as carbon (C) does not dramatically decrease as in the $W^{44+}$ ions. The total emission intensity decreases about 8% from the Absolute Extreme Ultraviolet (AXUV) photodiodes,[21] indicating that the high-Z impurities have the main contribution to the decrease of plasma radiation. It is also demonstrated that the ECRH makes main contribution to the core impurity suppression in this discharge.

## V. Conclusions

This study aims to absolutely calibrate TXCS on EAST by bremsstrahlung and obtain the $W^{44+}$ density profile on the basis of the W XLV (3.9095 Å) spectral line identified by TXCS. This study resulted in the calibration coefficient ~ $3.62 \times 10^{10}$ photons·m$^{-3}$·Å$^{-1}$·s$^{-1}$·sr$^{-1}$/(counts·s$^{-1}$·Å$^{-1}$) . This in-situ calibration method can be conducted without extra light source. Subsequently, the non-circular cross-section Abel inversion is employed to obtain the $W^{44+}$ density profile. Finally, the density profiles of $W^{44+}$ ions are compared with and without ECRH. The more flattened distribution is observed after ECRH injection which clearly demonstrates the impurity suppression effect of ECRH. It is also found that the total density of $W^{44+}$ is 4-5 orders of magnitude smaller than electron densities in typical EAST discharges. In the future work, the radial profile measurement by the absolutely calibrated XCS system is expected to be used to study W impurity transport in the core region.

## Acknowledgments

The authors wish to thank the team of EAST. The work is partially supported by National Magnetic Confinement Fusion Science Program of China (2017YFE0301300, 2019YFE03030002 and 2018YFE0303103), ASIPP Science and Research Grant (DSJJ-2020-02), Anhui Provincial Natural Science Foundation (NO.1908085J01, 2008085QA39), Comprehensive Research Facility for Fusion Technology Program of China (2018-000052-73-01-001228), Major science and technology infrastructure maintenance and reconstruction projects of Chinese Academy of Sciences (2021), the University Synergy Innovation Program of Anhui Province (No. GXXT-2021-029), The National Natural Science Foundation of China (12205072).

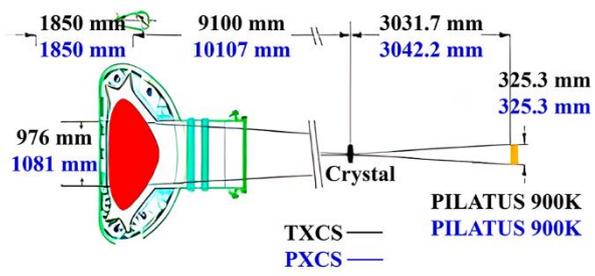

Fig. 1 The cross section of XCS system on EAST.

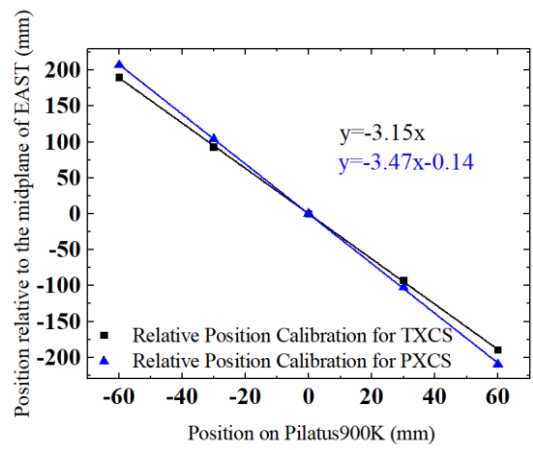

Fig.2 The position calibrations of XCS between the PILATUS 900K detector and the vertical position of EAST

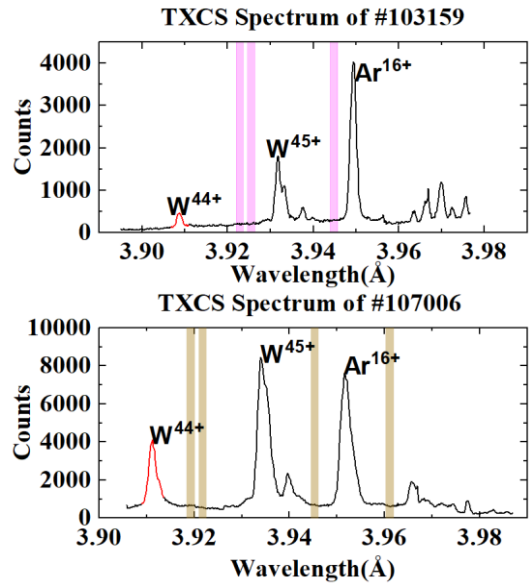

Fig.3 The typical TXCS spectra of EAST #103159 and #107006. The area blocks with color are the wavelength range chosen to analyze the bremsstrahlung.

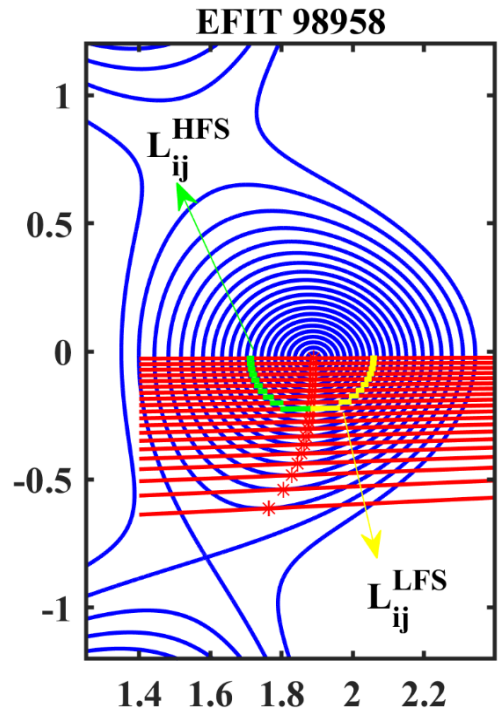

Fig.4 The principle of Abel inversion. The blue lines are the magnetic surface information obtained from EFIT, the red lines are the observation chords, the yellow lines are the $L_{ij}^{LFS}$, the green lines are the $L_{ij}^{HFS}$.

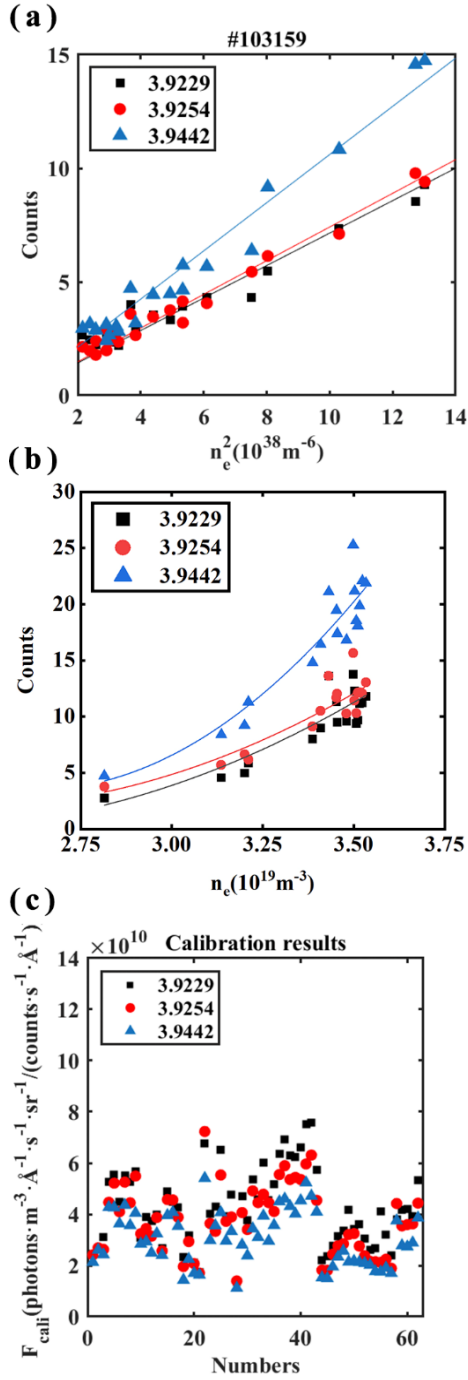

Fig. 5 (a) The relationship between the square of electron density with inversed detector counts of #103159. (b) The relationship between the electron density with inversed detector counts of several shots whose temperature are around 4.9 keV and $Z_{eff}$ are around 1.7. (c) The calibration coefficient results which concentrated distribution around the $4\times10^{10}$ photons·m$^{-3}$·Å$^{-1}$·s$^{-1}$·sr$^{-1}$/(counts·s$^{-1}$·Å$^{-1}$)

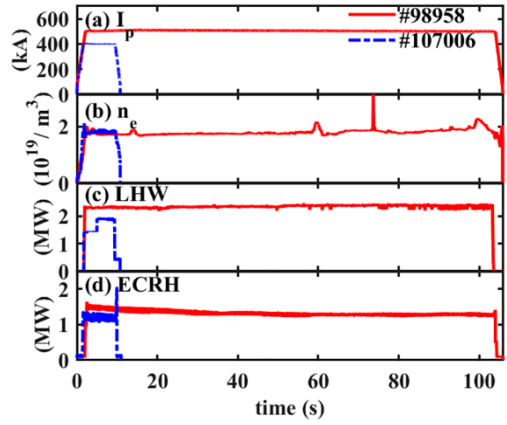

Fig. 6 The time evolution of (a)The plasma current (b) The electron density (c) The power of LHW (d) The power of ECRH.

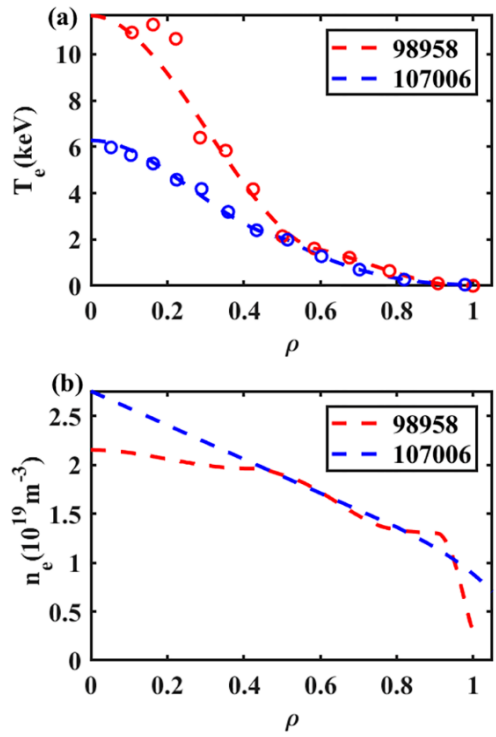

Fig. 7 (a) The electron temperature profile (b) The electron density profile of #98958 and #107006

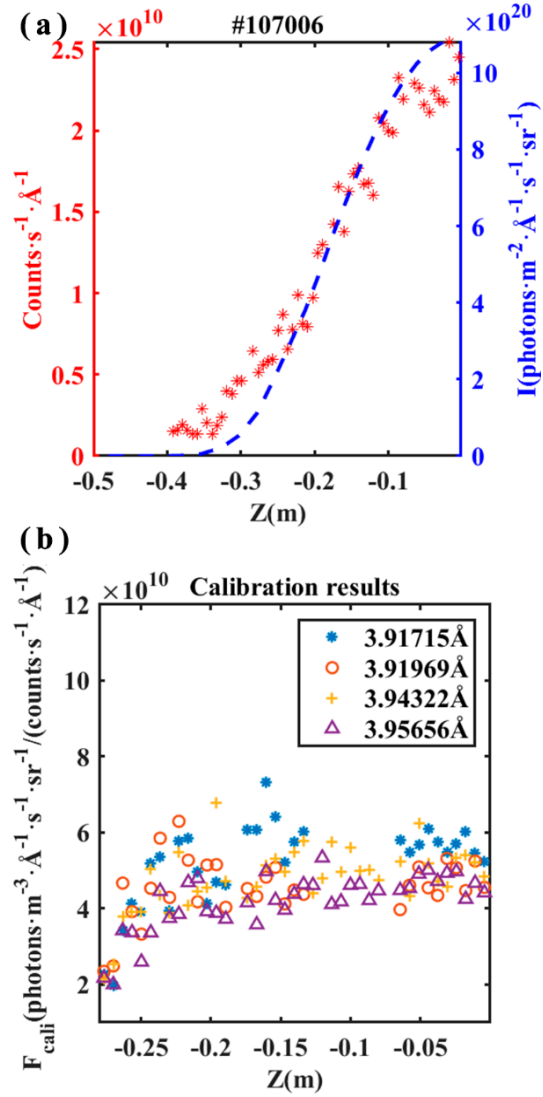

Fig. 8 (a)The calibration by comparing the line-integrated bremsstrahlung profile with measured counts (b)The calculation coefficient of EAST #107006

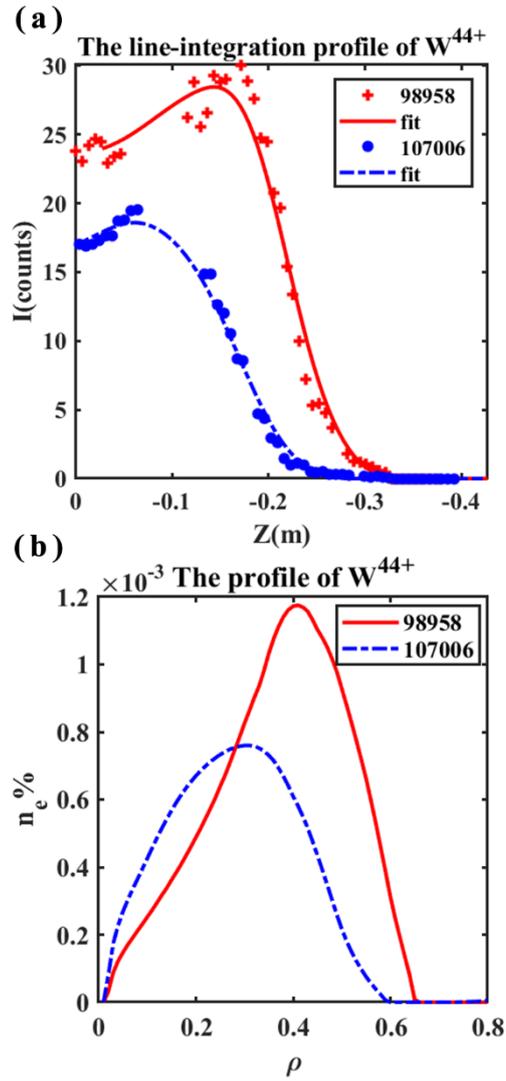

Fig. 9 (a) The experiment observed profiles and fitted curves of different electron temperature from EAST #98958 and #107006. (b) The $W^{44+}$ density profiles after the Abel inversion, which show a different distribution because of the electron temperature change

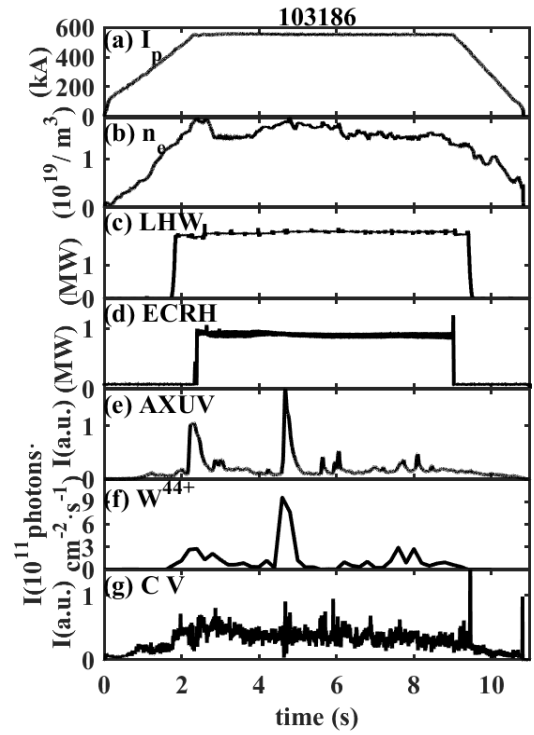

Fig. 10 The time evolution of (a) plasma current (b) electron density (c) the power of LHW (d) the power of ECRH (e) the total radiation power (f) absolute intensity of $W^{44+}$ by TXCS (g) C V by EUV

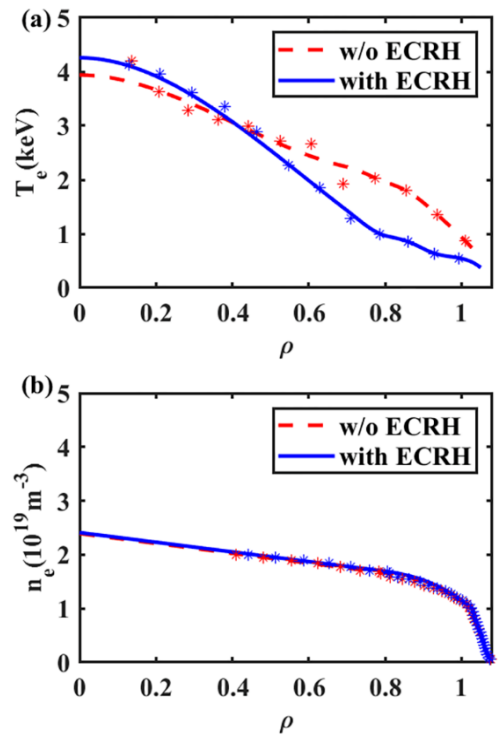

Fig. 11 (a) The profile of electron temperature measured by ECE (b) The profile of electron density measured by reflectometer and the core region is fitted by linear method.

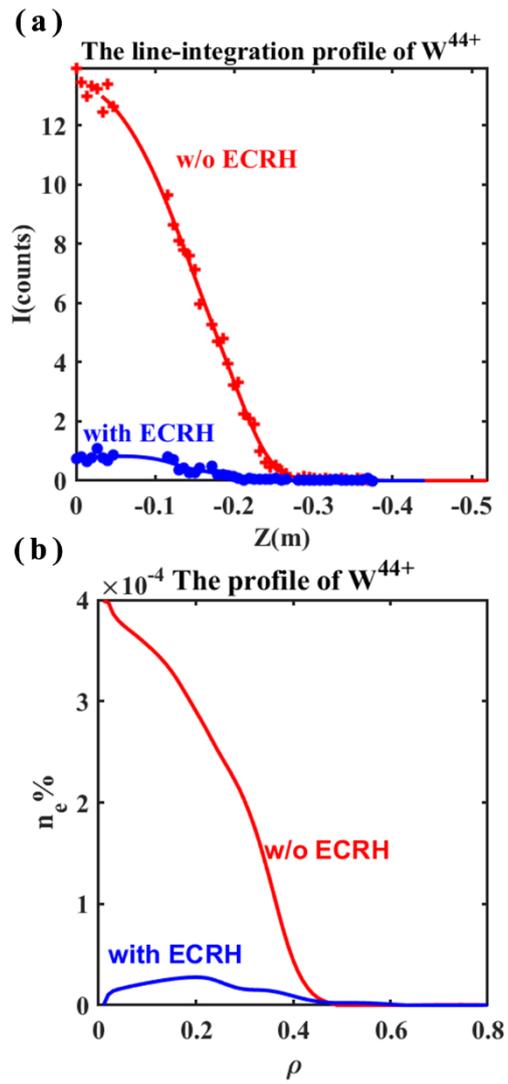

Fig. 12 (a)The comparation of the line-integration profiles between with and without ECRH (b) The density profile of $W^{44+}$ shows a dramatic intensity decrease after the ECRH applied